\documentclass[11pt]{article}
\usepackage{moriond,epsfig}
\usepackage[usenames]{color} 

\newcommand{\half}[0]{\frac{1}{2}}

\newcommand{\ld}[0]{\mathcal{L}}

\newcommand{\dd}[0]{{\rm{d}}}
\newcommand{\defn}[0]{\equiv}

\newcommand{\qsubrm}[2]{{#1}_{\scriptsize{\textrm{#2}}}}

\newcommand{\ep}[0]{{ {\delta}_{\scriptscriptstyle{\rm{E}}}}}
\newcommand{\lp}[0]{{ {\delta}_{\scriptscriptstyle{\rm{L}}}}}

 \newcommand{\kin}[0]{{\mathcal{X}}}
 \newcommand{\sol}[0]{\ld_{\scriptscriptstyle\{2\}}}

\newcommand{\lied}[1]{\pounds_{#1}}

\def\be{\begin{equation}}
\def\ee{\end{equation}}
\def\bea{\begin{eqnarray}}
\def\eea{\end{eqnarray}}

\begin{document}
\title{Effective field theory for perturbations in dark energy and modified gravity}

\author{Jonathan A. Pearson}

\address{Jodrell Bank Centre for Astrophysics, School of Physics and Astronomy,\\ The University of Manchester, M13 9PL}

\maketitle\abstracts{
When recent observational evidence and the GR+FRW+CDM model are combined we obtain the result that the Universe is accelerating, where the acceleration is due to some not-yet-understood ``dark sector". There has been a considerable number of theoretical models constructed in an attempt to provide an ``understanding" of the dark sector: dark energy and modified gravity theories. The proliferation of modified gravity and dark energy models has brought to light the need to construct a ``generic" way to parameterize the dark sector. We will discuss our new way of approaching this problem. We write down an effective action for linearized perturbations to the gravitational field equations for a given field content; crucially, our formalism does not require a Lagrangian to be presented for calculations to be performed and observational predictions to be extracted. Our approach is inspired by that taken in particle physics, where the most general modifications to the standard model are written down for a given field content that is compatible with some assumed symmetry (which we take to be isotropy of the background spatial sections).
}

\section{Introduction}
The standard cosmological model uses General Relativity (GR) as the gravitational theory, an FRW metric as the background geometry and a matter content of photons, baryons and cold dark matter. When this cosmological model is confronted with observational data of the Universe, a huge inconsistency arises: the observed acceleration\cite{sn} of the Universe is completely incompatible with the standard cosmological model. The standard ``fix'' for this is to include some mysterious substance into the matter content of the Universe, dark energy. One of the other ``fixes'' is to change the underlying gravitational theory for the Universe: perhaps GR is not the appropriate gravitational theory we should be applying on cosmological scales. Whether it is a modification to the gravitational theory or the addition of some dark energy, it is becoming clear that some as-yet unknown \textit{dark sector} must be introduced into the cosmological model. In the literature there are a plethora of modified gravity and dark energy theories whose aim it is to provide some description for the dark sector: $\Lambda$, quintessence, $k$-essence, galileon, Horndeski, TeVeS, \ae ther, $F(R)$, Gauss-Bonnet, Brans-Dicke are all such examples, to name but a few. The interested reader is directed to the extensive review\cite{clifton_mg}.

All modified gravity and dark energy theories have  gravitational field equations which can be written as
\bea
\label{eq:Sec:gen-grav0fes}
G_{\mu\nu} = 8 \pi G T_{\mu\nu} + U_{\mu\nu}
\eea
where $G_{\mu\nu}$ is the Einstein curvature tensor and contains the metric and its derivatives, $T_{\mu\nu}$ is the energy-momentum tensor for the known matter species in the Universe: it contains information about the content of the known sector. The final term, $U_{\mu\nu}$, is the \textit{dark energy-momentum tensor}, and parameterizes everything in the field equations that represents a deviation from the General Relativity + standard model particles picture.   Equation (\ref{eq:Sec:gen-grav0fes}) can be split up using perturbation theory: the unperturbed term provides an   equation governing the time evolution of the scale factor, where the energy density and pressure of the dark sector fluid are given by $\qsubrm{\rho}{dark} = {U^0}_0, \qsubrm{P}{dark} = \frac{1}{3}{U^i}_i$.
To linearized perturbations, the equation contains information governing the evolution of structures in the Universe,
\bea
\delta G^{\mu\nu} = 8 \pi G \delta T^{\mu\nu} + \delta U^{\mu\nu}.
\eea
The quantity $\delta U^{\mu\nu}$ is the \textit{perturbed dark energy momentum tensor} and parameterizes the deviation of the gravitational theory from GR at perturbed order. The salient question is, therefore, ``\textit{how do we write down the consistent and physically meaningful  deviations from GR?}'' How do we construct the allowed set of $\delta U^{\mu\nu}$? This has been studied by a number of authors, see e.g.  \cite{bakeretal}.

Of course, known theories will provide prescriptions for what the $\delta U^{\mu\nu}$ could be. However, we would like a tractable approach   which does not require a theory to be presented at all. Our approach \cite{batpears} is to construct a Lagrangian   for the perturbations, and use that to construct the perturbed dark energy-momentum tensor. The advantage of using a Lagrangian is that all the usual field theoretic techniques can be employed, and all freedom in the theory can be traced back to some interaction term in the Lagrangian. This is the same approach taken in particle physics. For example: the most general modifications to the standard model potential are written down under a very small number of assumptions (usually just a field content and some symmetries),  all ``new'' mass and interaction terms are identified and then experiments are devised to constrain the values of these new terms.

We are able to identify the maximum number of free functions under a very general set of theoretical priors we impose on the theory (such as the field content). The number of free functions   decreases as soon as extra theoretical priors are imposed (such as symmetries of the background spacetime).

\section{The Lagrangian for perturbations: formalism}
We will consider a dark sector which is constructed from some set of  field variables $\{X^{\scriptscriptstyle(\rm{A})}\}$; this includes the metric, vector and scalar fields. In perturbation theory each of the field variables is written as a perturbation about some homogeneous background value: $
X^{\scriptscriptstyle(\rm{A})} = \bar{X}^{\scriptscriptstyle(\rm{A})} + \delta X^{\scriptscriptstyle(\rm{A})}$.
 We will construct the   Lagrangian  for perturbations from \textit{Lagrangian} perturbed field variables, denoted by $\lp X^{\scriptscriptstyle(\rm{A})}$, and these are given in terms of the \textit{Eulerian} perturbation of the field variable, $\ep X^{\scriptscriptstyle(\rm{A})}$, via the Lie derivative of the field variable along some diffeomorphism generating vector field, $\xi^{\mu}$, according to
\bea
\lp X^{\scriptscriptstyle(\rm{A})} = \ep X^{\scriptscriptstyle(\rm{A})}  + \lied{\xi} X^{\scriptscriptstyle(\rm{A})}.
\eea
We must   be provided with information as to whether the field variables $X^{\scriptscriptstyle(\rm{A})}$ are scalars, vectors, tensors etc, so that we can incorporate the correct combinations of the $\xi^{\mu}$-field in the Lagrangian. For example, one can   compute the relationship between the Eulerian and Lagrangian perturbations to the metric: $\lp g_{\mu\nu} = \ep g_{\mu\nu} + 2 \nabla_{(\mu}\xi_{\nu)}$.

The   Lagrangian   which will give linearized field equations in the perturbed field variables is a  quadratic functional of the Lagrangian perturbed field variables. The   Lagrangian for perturbations, $\sol$,  is equivalent to the second measure-weighted variation of the action:
\bea
\delta^2S = \int \dd^4x\sqrt{-g}\,\bigg\{\frac{1}{\sqrt{-g}}\delta^2(\sqrt{-g}\ld)\bigg\} = \int \dd^4x\sqrt{-g}\, \sol.
\eea
The   Lagrangian for perturbations will contribute towards the gravitational field equations via a perturbed energy momentum tensor, which is a linear functional in the Lagrangian perturbed field variables, $\lp U^{\mu\nu} = \lp U^{\mu\nu}[\lp X^{\scriptscriptstyle(\rm A)}]$, and is computed via
\bea
\lp U^{\mu\nu} = - \half \bigg[ 4\frac{\hat{\delta}}{\hat{\delta} \lp g_{\mu\nu}}\sol+ U^{\mu\nu}g^{\alpha\beta}\lp g_{\alpha\beta}\bigg].
\eea
The perturbed dark energy momentum tensor also satisfies the perturbed conservation equation, $\delta(\nabla_{\mu}U^{\mu\nu})=0$. The perturbations relevant for cosmology are the Eulerian perturbations (because they are performed around   the FRW background geometry).

\section{Examples}
We will now provide two  examples. To begin, the simplest possible example, where the dark sector does not contain anything extra: only the metric is present. Hence, we are considering a dark sector theory with field content given by
\bea
\ld = \ld(g_{\mu\nu}).
\eea
The   Lagrangian   for perturbations is given by the quadratic functional
\bea
\sol=\frac{1}{8} \mathcal{W}^{\mu\nu\alpha\beta} \lp g_{\mu\nu}\lp g_{\alpha\beta}.
\eea
We had to introduce one rank-4 tensor, $\mathcal{W}$, which is only a function of background field variables (for an FRW background, $\mathcal{W}$ is   specified in terms of 5 time   dependent functions). The perturbed dark energy momentum tensor is given by
\bea
\lp U^{\mu\nu} = - \half \bigg\{ \mathcal{W}^{\mu\nu\alpha\beta} + U^{\mu\nu}g^{\alpha\beta}\bigg\}\lp g_{\alpha\beta}.
\eea
This  encompasses elastic dark energy and massive gravity models in GR, such as the Fierz-Pauli theory. As a second example, we consider a dark sector containing the metric, a scalar field $\phi$ and its first derivative
\bea
\ld = \ld(g_{\mu\nu}, \phi, \nabla_{\mu}\phi).
\eea
In this case, the   Lagrangian for perturbations is given by the quadratic functional
\bea
\sol  &=& \mathcal{A} (\delta\phi)^2+\mathcal{B}^{\mu}   \delta\phi \nabla_{\mu}  \delta\phi   +\half\mathcal{C}^{\mu\nu}\nabla_{\mu} \delta\phi \nabla_{\nu}  \delta\phi \nonumber\\
&&\qquad+ \frac{1}{4}\bigg[ \mathcal{Y}^{\alpha\mu\nu}   \nabla_{\alpha} \delta\phi \lp g_{\mu\nu}+ \mathcal{V}^{\mu\nu}  \delta\phi \lp g_{\mu\nu} +\half\mathcal{W}^{\mu\nu\alpha\beta}  \lp g_{\mu\nu}\lp g_{ \alpha\beta}\bigg] .  
\eea
We had to introduce 6 tensors, each of which is only a function of background field variables. These tensors describe the interactions of the field content in the Lagrangian. The perturbed dark energy momentum tensor is given by
\bea
\lp U^{\mu\nu} =  -\half \bigg\{   \mathcal{V}^{\mu\nu} \delta\phi +  \mathcal{Y}^{\alpha\mu\nu} \nabla_{\alpha}  \delta\phi \bigg\}    -\half \bigg\{\mathcal{W}^{\alpha\beta\mu\nu}  +  g^{\alpha\beta} U^{\mu\nu} \bigg\}\lp g_{\alpha\beta}.
\eea
This model is similar to that studied in \cite{creminelli}.
This encompasses quintessence, $k$-essence, and Lorentz violating theories. Notice that the generalized gravitational field equations at perturbed order will be entirely specified once the components of the three tensors $\mathcal{V}, \mathcal{Y}, \mathcal{W}$ are specified.   If we do not impose any symmetries or theoretical structure upon the theory, then there are 74 free functions in these three tensors. As soon as we impose spatial isotropy of the background one finds that there are  10 free functions, and imposing the theoretical structure: $\ld = \ld(\phi, \kin)$ where $ \kin \defn - \half g^{\mu\nu}\nabla_{\mu}\phi\nabla_{\nu}\phi$ is the kinetic scalar, there are only 3 free functions. These three functions can conceivably be confronted with experimental data to constrain  the possible values that they can take.

\section{Cosmological perturbations and entropy}
If we impose isotropy upon the spatial sections of the background we are able to split the background tensors which appear in $\sol$ using an isotropic (3+1) decomposition using a time-like unit vector $u^{\mu}$ and a 3D metric $\gamma_{\mu\nu}$. This means that all tensors are immediately compatible with an FRW background,  the relevant equations considerably simplify and the number of free functions dramatically decreases (each free function is now only a function of time).

The Eulerian perturbed dark energy-momentum tensor can be written as a fluid decomposition
\bea
\ep {U^{\mu}}_{\nu} = (\delta\rho + \delta P) u^{\mu}u_{\nu} + \delta P {\delta^{\mu}}_{\nu} + (\rho+P)(v^{\mu}u_{\nu} + u^{\mu}v_{\nu}) + P{\Pi^{\mu}}_{\nu},
\eea
where $\rho,P$ are the density and pressure of the dark sector fluid and ${\Pi^{\mu}}_{\nu}$ is the anisotropic sources ($u^{\mu}$ is the time-like unit vector and $v^{\mu}$ is a space-like vector). The entropy perturbation $w\Gamma\defn ( {\delta P}/{\delta\rho} -  {\dd P}/{\dd\rho})\delta $, can also be identified.  The perturbed conservation equations provide evolution equations for $\delta, v$ but not $\delta P, \Pi$; thus, once $\delta P, \Pi$ are provided by some means, the system of equations becomes closed and can be solved. Our formalism enables us to obtain general forms of $\Gamma,\Pi$. For a scalar field theory of the type $\ld = \ld(\phi, \kin)$, where $\kin \defn - \half g^{\mu\nu}\nabla_{\mu}\phi\nabla_{\nu}\phi$ is the kinetic term of the scalar field (and can be thought of as a theoretical prior on how the metric and derivative of the scalar field combine in the theory) one finds that $\Pi =0$ identically and the entropy perturbation can be written in terms of \textit{two free functions of time}:
\bea
\label{eq:sec;entropy-proc-jp}
w\Gamma = ({ { }{\alpha}} - w) \bigg[ \delta - 3 \mathcal{H}{ { }{\beta}}(1+w)\theta\bigg].
\eea
This $(\alpha,\beta)$-parameterization of the entropy perturbation (\ref{eq:sec;entropy-proc-jp}) is more general than that given in  \cite{wellerlewis}, who had $\beta=1$.
For a general kinetic scalar field theory, one finds
\bea
{ { }{\alpha}} = \bigg( 1 + 2 \kin \frac{\ld_{,\kin\kin}}{\ld_{,\kin}}\bigg)^{-1},\qquad { { }{\beta}} = \frac{2a\ld_{,\phi}}{3\mathcal{H}\ld_{,\kin}\sqrt{2\kin}} \bigg[ 1 + \kin \bigg( \frac{\ld_{,\kin\kin}}{\ld_{,\kin}} - \frac{\ld_{,\kin\phi}}{\ld_{,\phi}}\bigg) \bigg] \frac{{ { }{\alpha}}}{{ { }{\alpha}}-w}.
\eea
In quintessence models, $\ld = \kin - V(\phi)$, one finds that $({ { }{\alpha}}, { { }{\beta}})= (1,1)$ and in pure $k$-essence models, $\ld = \ld(\kin)$, one finds that $({ { }{\alpha}}, { { }{\beta}})= ({ { }{\alpha}},0)$. 

\section{Discussion}
In this paper we have merely given a flavour of our formalism \cite{batpears} and how we use it to  construct the generalized perturbed gravitational field equations. The formalism provides a systematic way to isolate all the freedom within wide classes of models, and to obtain an understanding how this freedom translates into cosmologically observable quantities (such as the CMB, matter power and lensing spectra). At this stage, the most useful result of our formalism is given by equation (\ref{eq:sec;entropy-proc-jp}).

 In future work we will show how our formalism can be extended for use with theories with high derivatives (relevant  for  theories containing curvature tensors or galileons). We will also show what the observational signatures   of our generalized theories are, where we will use current datasets to provide constraints on the free parameters.
 
This work was done in collaboration with R.A. Battye, and we acknowledge stimulating conversations with T. Baker, P. Ferreira, A. Lewis and R. Bean.

\end{document}